\documentclass[conference,letterpaper]{IEEEtran}
\IEEEoverridecommandlockouts
\usepackage{xcolor}

\usepackage[utf8]{inputenc}
\usepackage[T1]{fontenc}
\usepackage{url}
\usepackage{ifthen}
\usepackage{cite}
\usepackage[cmex10]{amsmath} 
\usepackage{amssymb,amsfonts}
\usepackage{algorithmic}
\usepackage{graphicx}
\usepackage{textcomp}
\usepackage{xcolor}
\usepackage{tikz}
\allowdisplaybreaks

\usepackage{afterpage}

\newcommand{\msf}[1]{\mathsf{#1}}

\makeatletter
\def\@begintheorem#1#2{%
  \trivlist
  \item[\hskip \labelsep{\bfseries #1\ #2.}]\normalfont
}
\def\@opargbegintheorem#1#2#3{%
  \trivlist
  \item[\hskip \labelsep{\bfseries #1\ #2\ {\normalfont(#3)}.}]\normalfont
}
\makeatother

\newtheorem{lemma}{Lemma}
\newtheorem{remark}{Remark}

\newtheorem{proposition}{Proposition}
\newtheorem{corollary}{Corollary}

\def\BibTeX{{\rm B\kern-.05em{\sc i\kern-.025em b}\kern-.08em
    T\kern-.1667em\lower.7ex\hbox{E}\kern-.125emX}}

\begin{document}

\title{Design of APSK Constellations Approaching the Communication-Sensing Pareto Boundary for ISAC
}

\author{
\IEEEauthorblockN{
Yujie Shao$^\dagger$, Min Qiu$^\dagger$\thanks{Corresponding author: Min Qiu.},
Ming-Chun Lee$^*$, Yu-Chih Huang$^*$, and Jinhong Yuan$^\ddagger$
}
\IEEEauthorblockA{
$^\dagger$Global College, Shanghai Jiao Tong University, Shanghai 200240, China\\
$^*$Institute of Communications Engineering, National Yang Ming Chiao Tung University, Hsinchu 300, Taiwan\\
$^\ddagger$School of Electrical Engineering and Telecommunications, University of New South Wales, Sydney, NSW 2052, Australia\\
E-mail: \{shaox3, min\_qiu\}@sjtu.edu.cn, \{mingchunlee, jerryhuang\}@nycu.edu.tw, \{j.yuan\}@unsw.edu.au
}
}

\maketitle

\begin{abstract}
We propose a semi‑analytical design framework for amplitude phase shift keying (APSK) signaling 
for integrated sensing and communication (ISAC), focusing on i.i.d. uniform discrete input distributions 
for practicality and analytical tractability.
First, we establish APSK design criteria in which communication performance is measured by the gap to 
capacity and linked to the minimum Euclidean distance, while sensing performance is characterized by the 
symbol‑energy variance.
Based on these criteria, we develop a structured APSK construction whose key parameters follow 
explicit scaling laws. Then we prove that this construction achieves a constant gap to capacity \emph{independent} 
of the signal‑to‑noise ratio. Building upon this foundation, we further construct a parametric 
APSK family that bridges the communication‑optimal and sensing‑optimal designs, with the 
communication and sensing (C\&S) tradeoff controlled by the number of rings and energy allocation among 
rings. 
Simulation results show that the resulting APSK family achieves C\&S performance very close to 
the Pareto boundary achieved with time‑independent, circularly symmetric, and otherwise \emph{unconstrained} 
continuous input distributions.
\end{abstract}

\begin{IEEEkeywords}
ISAC, APSK, shaping.
\end{IEEEkeywords}

\section{Introduction}
Integrated sensing and communication (ISAC) has been envisioned as a key enabling technology for 6G networks, where wireless systems are expected to provide both high-rate communication and environment-aware sensing capabilities~\cite{Fanliu2022JSAC,Zhang2022,AnLiu2022}. By reusing spectrum, hardware, and waveform resources, ISAC improves spectrum utilization and overall system efficiency compared with separately designed sensing and communication systems, while reducing hardware cost and deployment complexity and facilitating mutual enhancement between sensing and communication through shared waveforms and environmental information~\cite{AnLiu2022}.

Due to the distinct requirements of communication and sensing (C\&S), ISAC systems inherently face a tradeoff between C\&S performance. 
Motivated by the promising potential of ISAC, existing studies have explored various design directions, including waveform design~\cite{Zhou2022,Gu2022}, beamforming design~\cite{FLiu2022TSP,Choi2024}, and network-level ISAC system design~\cite{Meng2024TWC,Meng2025WC}, to achieve more favorable C\&S tradeoffs. 
Meanwhile, the fundamental limits of the C\&S tradeoff have also been investigated from information-theoretic and network-level perspectives~\cite{AnLiu2022,Xiong2023,Qiu2026}.

To fully exploit the potential of ISAC with a shared waveform, the design of dual-functional signals that approach the optimal C\&S tradeoff is essential.
In~\cite{Guo2026TWC}, the authors proposed a modified Blahut–Arimoto algorithm (MBA) to numerically obtain the Pareto boundary of the Cram\'er--Rao bound (CRB)–Rate region by calculating the optimal boundary-achieving distributions that are time-independent and circularly symmetric. However, these input distributions are difficult to implement in practical systems.
On the other hand, several works have considered discrete constellations and designed geometric and/or probabilistic shaping schemes for orthogonal frequency division multiplexing (OFDM)-ISAC systems to achieve improved C\&S tradeoff, e.g., between mutual information (MI) and sensing ambiguity-function sidelobes~\cite{Du2024}, and between MI and detection probability~\cite{Geiger2026}. Very recently, Gamma-distributed geometric constellation design has also been proposed for single-carrier ISAC to balance mutual information and detection probability~\cite{keshavarzchafjiri2026}.
However, the above designs are based on numerical optimization because the mutual information of finite-alphabet inputs generally lacks a tractable closed-form expression and is evaluated by the Monte Carlo simulation. 
In addition, the exact optimal tradeoff between C\&S is not available in closed form, making the derivation of the optimal input distribution for ISAC more challenging.

In this paper, we make a first step toward the semi-analytical design of discrete input distributions for ISAC.
We consider i.i.d. inputs uniformly distributed over a finite discrete constellation. 
This setting is consistent with practical standardized constellations and avoids the additional transceiver complexity that may arise from probabilistic shaping. 
In particular, we adopt amplitude phase shift keying (APSK) signaling, 
which not only offers a flexible structure and a large design space, but is also used in existing communication standards, e.g., satellite communications~\cite{ETSI_EN_302_307_2_2024}.

The main contributions of this paper are summarized as follows.
First, we establish a design criterion for APSK signaling, 
where communication performance is measured by the gap to capacity, while sensing performance is quantified by the symbol-energy variance. Both metrics admit analytically tractable expressions.
We then propose a semi-analytical design framework for APSK, 
whose key design parameters, including the number of rings, the ring radii, and the number of constellation points on each ring, follow explicit scaling laws.
We prove that this structured design achieves a constant gap to capacity regardless of the signal-to-noise ratio (SNR), yielding asymptotically communication-optimal performance.
On the sensing side, the optimal design is constant-modulus signaling.
Building upon the communication-optimal and sensing-optimal designs, we propose a semi-analytical parametric family of APSK that bridges these two extremes. The constellation structure is analytically prescribed except for a small number of parameters for numerical tuning. 
The resulting C\&S tradeoff can be controlled by the number of rings and energy allocation among rings.
Simulation results show that the proposed APSK family achieves a C\&S tradeoff very close to the Pareto boundary of the symbol-energy variance-rate region generated by MBA with time-independent, circularly symmetric, and otherwise unconstrained continuous input distributions~\cite{Guo2026TWC}. In addition, the proposed schemes significantly outperform the time-sharing baseline scheme.

\emph{Notations}: 
The sets of natural numbers and complex numbers are presented by $\mathbb{N}$ and $\mathbb{C}$, respectively.
Random variables are written in upper-case sans serif font, e.g., $\msf{X}$. 
Scalar realizations are written in lower-case letters while 
vectors are written in bold lower-case letters.
The notations $\mathcal{O}(\cdot)$, $\Theta(\cdot)$, and $\Omega(\cdot)$ denote asymptotic upper, tight, and lower bounds, respectively.
For an angle $\phi$, $|\phi|_{2\pi}\triangleq \min_{m\in\mathbb{Z}}|\phi+2\pi m|$.

\section{System Model}
We consider a monostatic ISAC system, where an ISAC base station (BS) serves a communication user while simultaneously performing sensing tasks using the reflected transmitted signals.

The BS generates a dual-functional waveform sequence
$\boldsymbol{x} = [x_1, \cdots, x_L]\in \mathbb{C}^L,$
where $L$ denotes the symbol length. 
The symbols $\{x_\ell\}_{\ell=1}^L$ 
are assumed to be i.i.d. with each $x_{\ell}$ uniformly distributed over a discrete constellation set $\mathcal{X}$, which is consistent with practical digital modulation schemes. The constellation is normalized such that $\mathbb{E}[|x_\ell|^2] = 1$. The transmitted signals are $\sqrt{P}\boldsymbol{x}$, where $P$ denotes transmit power.

\subsection{Communication Channel Model}
The received signal at the user is given by
\begin{align}
    \boldsymbol{y}_{\text{c}} = \sqrt{P}h_{\text{c}}\boldsymbol{x} + \boldsymbol{z}_{\text{c}}\in \mathbb{C}^L,
\end{align}
where $h_{\text{c}}\in \mathbb{C}$ is the communication channel coefficient, and $\boldsymbol{z}_{\text{c}} \sim \mathcal{CN}(\boldsymbol{0}, \sigma_{\text{c}}^2\boldsymbol{I}_L)$ denotes the additive Gaussian noise. We assume quasi-static channel fading and perfect CSI at the user, i.e., $h_{\text{c}}$ is known at the receiver. The communication SNR is defined as $\mathrm{SNR}_{\text{c}} \triangleq \frac{P|h_{\text{c}}|^2}{\sigma_{\text{c}}^2}$.

For the communication subsystem, we adopt the mutual information as the performance metric.
Since the symbols $\{x_\ell\}_{\ell=1}^{L}$ are i.i.d., we consider a generic
transmitted symbol $\msf{X}$, with realization $x$, and the corresponding received
symbol $\msf{Y}_{\text{c}}$, with realization $y_{\text{c}}$. For a given channel
realization $h_{\text{c}}$, the per-symbol mutual information can be written as~\cite{2006Elements}
\begin{equation}
    I(\msf{X};\msf{Y}_{\text{c}})
    =
    \mathbb{E}_{\msf{X},\msf{Y}_{\text{c}}}
    \left[
    \log_2
    \frac{
    \mathbb{P}_{\msf{Y}_{\text{c}}|\msf{X}}(y_{\text{c}}|x)
    }{
    \sum_{x_i \in \mathcal{X}} \mathbb{P}_{\msf{X}}(x_i)
    \mathbb{P}_{\msf{Y}_{\text{c}}|\msf{X}}(y_{\text{c}}|x_i)
    }
    \right].
    \label{2}
\end{equation}
where
\begin{equation}
    \mathbb{P}_{\msf{Y}_{\text{c}}|\msf{X}}(y_{\text{c}}|x)
    =
    \frac{1}{\pi\sigma_{\text{c}}^{2}}
    \exp\!\left(
    -\frac{|y_{\text{c}}-\sqrt{P}h_{\text{c}}x|^2}
    {\sigma_{\text{c}}^2}
    \right).
\end{equation}
Accordingly, the achievable rate per channel use is $R = I(\msf{X};\msf{Y}_{\text{c}}).$

\subsection{Sensing Channel Model}

The reflected signal received at the BS is modeled as
\begin{align}
    \boldsymbol{y}_{\text{s}}
    =
    \sqrt{P}h_{\text{s}}\boldsymbol{x}
    +
    \boldsymbol{z}_{\text{s}}\in \mathbb{C}^L,
\end{align}
where $h_{\text{s}} \in \mathbb{C}$ is an unknown deterministic sensing channel coefficient to be estimated, assumed to be quasi-static over the observation block of length $L$, and
$\boldsymbol{z}_{\text{s}}\sim\mathcal{CN}(\boldsymbol{0},\sigma_{\text{s}}^2\boldsymbol I_L)$
denotes the sensing noise.

For the sensing subsystem, we characterize the sensing performance by the
classical conditional CRB for estimating $h_{\text{s}}$ given $\boldsymbol{x}$.
According to~\cite[Eq. (21)]{keshavarzchafjiri2026},
the conditional CRB is
\begin{equation}
    \mathrm{CRB}_{h_{\text{s}}}(\boldsymbol{x})
    =
    \frac{\sigma_{\text{s}}^2}
    {P\|\boldsymbol{x}\|^2}.
\end{equation}

As a scalar sensing metric, we consider the average conditional CRB:
\begin{equation}
    \overline{\mathrm{CRB}}_{h_{\text{s}}}
    =
    \mathbb E_{\boldsymbol x}
    \!\left[
    \mathrm{CRB}_{h_{\text{s}}}(\boldsymbol{x})
    \right]
    =
    \frac{\sigma_{\text{s}}^2}{P}
    \mathbb E_{\boldsymbol x}
    \!\left[
    \frac{1}
    {\sum_{\ell=1}^{L}|x_\ell|^2}
    \right].
    \label{eq:avg_CRB_exact}
\end{equation}
The metrics $R$ and $\overline{\mathrm{CRB}}_{h_{\text{s}}}$ are jointly used to characterize the C\&S tradeoff.

\section{APSK Signaling for ISAC}
\begin{figure}[t]
\centering
\begin{tikzpicture}[scale=0.8]

\draw[dashed, gray] (0,0) circle (1.0);
\draw[dashed, gray] (0,0) circle (1.8);
\draw[dashed, gray] (0,0) circle (2.6);

\foreach \a in {20,80,140,200,260,320}{
    \fill ({1.0*cos(\a)},{1.0*sin(\a)}) circle (2pt);
}

\foreach \a in {0,45,90,135,180,225,270,315}{
    \fill ({1.8*cos(\a)},{1.8*sin(\a)}) circle (2pt);
}

\foreach \a in {10,40,70,100,130,160,190,220,250,280,310,340}{
    \fill ({2.6*cos(\a)},{2.6*sin(\a)}) circle (2pt);
}

\draw[thick,dashed] (0,0) -- (1.8,0);

\node[right] at (1.0,0.8) {$k\text{-th ring}$};

\draw[thick,->] (0,0) -- ({1.8*cos(45)},{1.8*sin(45)});
\node[left] at (0.65,0.65) {$r_k$};

\draw[thick,purple] (0.5,0) arc[start angle=0,end angle=45,radius=0.5]
    node[pos=0.6, right,text=purple] {$\phi_k+\frac{2\pi n}{N_k}$};

\draw[thick,<->,blue] 
    ({1.8*cos(90)},{1.8*sin(90)}) -- 
    ({1.8*cos(45)},{1.8*sin(45)});
\node[right,text=blue] at (0.45,1.75) {$d_k^{\mathrm{intra}}$};

\draw[thick,<->,red] 
    ({1.8*cos(45)},{1.8*sin(45)}) -- 
    ({2.6*cos(40)},{2.6*sin(40)});
\node[right,text=red] at (2.0,1.5) {$d_{k,k'}^{\mathrm{inter}}$};

\end{tikzpicture}
\caption{Illustration of an APSK constellation.}
\label{fig:apsk}
\end{figure}
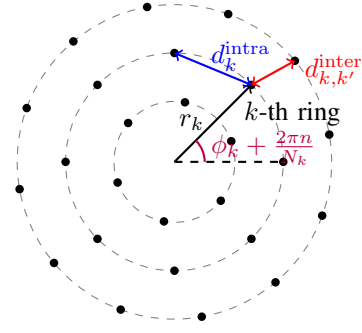

In this section, we present the APSK signaling considered here for ISAC and investigate the tradeoff between C\&S performance.

\subsection{Preliminaries}
Consider a $2^m$-APSK constellation $(m\ge2)$. Let \(K_m\) denote the number of rings, and let \(N_k\) and \(r_k\) denote the number of points and the radius of the \(k\)-th ring, respectively, where \(\sum_{k=1}^{K_m}N_k=2^m\). The APSK constellation is given by
\begin{align}
\mathcal{X}
=
\bigcup_{k=1}^{K_m}
\left\{
r_k e^{j\left(\phi_k+\frac{2\pi n}{N_k}\right)} : n=0,1,\dots,N_k-1
\right\},
\label{APSK}
\end{align}
where \(\phi_k\) is the phase offset of the \(k\)-th ring. 
The constellation is normalized such that the average symbol energy satisfies
$\frac{1}{2^m}\sum_{x\in\mathcal{X}} |x|^2 = 1$.
The minimum Euclidean distance of the APSK constellation \(\mathcal{X}\) is given by
\begin{align}
    d_{\min}(\mathcal{X}) = \min_{\substack{1\le k,k'\le K_m \\0\le n<N_k, 0\le n' <N_{k'}\\(k,n)\neq(k',n')}}
    \sqrt{r_k^2+r_{k'}^2-2r_kr_{k'}\cos\theta},
\end{align}
where $\theta=\phi_k-\phi_{k'}+\frac{2\pi n}{N_k}-\frac{2\pi n'}{N_{k'}}$.
As illustrated in Fig.~\ref{fig:apsk}, the minimum distance can be decomposed into intra-ring and inter-ring distances, with intra-ring and inter-ring distances
given by
\begin{subequations}\label{dmin_general_component}
\begin{align}
    &d_k^{\text{intra}}
    = 2r_k\sin\left(\frac{\pi}{N_k}\right), \label{dminintra}\\
    &d_{k,k'}^{\text{inter}}
    = \sqrt{r_k^2+r_{k'}^2-2r_kr_{k'}\cos\Delta_{k,k'}} \quad(k\neq k'),\label{dmininter}\\
    &\Delta_{k,k'}
    = \min_{\substack{0\le n<N_k\\0\le n'<N_{k'}}}
    \left|
    \phi_k-\phi_{k'}+\frac{2\pi n}{N_k}-\frac{2\pi n'}{N_{k'}}
    \right|_{2\pi}.
\end{align}
\end{subequations}
Therefore, the minimum distance of $\mathcal{X}$ can be expressed as
\begin{align}
\label{dmin_general}
    d_{\min}(\mathcal{X})
    =
    \min\left\{
    \min_{1\le k \le K_m} d_k^{\text{intra}},
    \;
    \min_{\substack{1\le k<k'\le K_m}} d_{k,k'}^{\text{inter}}
    \right\}.
\end{align}

\subsection{Design Criterion for Communication}

We evaluate the communication performance by analyzing the gap between the mutual information and capacity.
From~\eqref{2}, we have
\begin{subequations}\label{mutualbound}
\begin{align}
&I(\msf{X};\msf{Y}_{\text{c}})
= I(\sqrt{P}h_{\text{c}}\msf{X};
\sqrt{P}h_{\text{c}}\msf{X}+\msf{Z}_{\text{c}}), \label{mutualbounda}\\
&\ge H(\msf{X})
- \log_2\left(\frac{2\pi e}{4}\right)
- \log_2\left(
1+\frac{16}{\pi\mathrm{SNR}_{\text{c}}d_{\min}^2(\msf{X})}
\right), \label{mutualboundb}
\end{align}
\end{subequations}
where~\eqref{mutualboundb} follows by applying Lemma~\ref{mutualboundlemma} in Appendix A to lower bound the mutual information of any two-dimensional constellation.
Therefore, the gap between mutual information and the Gaussian capacity $C({\mathrm{SNR}_{\text{c}}})$ is upper bounded as
\begin{align}
    &C({\mathrm{SNR}_{\text{c}}})- I(\msf{X};\msf{Y}_{\text{c}}) < \log_2\!\left(\frac{2\pi e}{4}\right) \notag \\
    &+\log_2\!\left(\frac{1}{2^m}\Big(1+\mathrm{SNR}_{\text{c}}+\frac{16}{\pi d_{\min}^2(\msf{X})\mathrm{SNR}_{\text{c}}}+\frac{16}{\pi d_{\min}^2(\msf{X})}\Big)\right).
    \label{gap}
\end{align}
Following~\eqref{gap}, our design criterion for achieving the (asymptotic) communication-optimal point is to maintain a constant gap to capacity, \emph{independent of $\mathrm{SNR}_{\text{c}}$}.
\begin{remark}
{\itshape
A constant gap to capacity implies that the analytical upper bound on the capacity gap remains bounded independently of $\mathrm{SNR}_{\text{c}}$}. In other words, the proposed communication-optimal signaling does not need to be redesigned for each $\mathrm{SNR}_{\text{c}}$ value.\hfill$\blacksquare$
\end{remark}
From~\eqref{gap}, the key to achieving a constant gap lies in $d_{\min}(\msf{X})$. 
To quantify how well an APSK constellation can perform,
the following lemma provides an upper bound on the minimum distance of a general APSK constellation.
\begin{lemma}
{\itshape
Let $\msf{X}$ be uniformly distributed over a normalized $2^m$-APSK constellation $\mathcal{X}$ in~\eqref{APSK},
then there exists a positive constant $Q$ such that $d_{\min}^2(\msf{X}) \le \frac{Q}{2^m}.$
}
\label{dminenergy}
\end{lemma}
\IEEEproof{
The proof is in Appendix B.\hfill$\blacksquare$
}

In the next subsection, we present a semi-analytical design to achieve minimum distance upper bound in the same order.

\subsection{Asymptotically-Optimal Communication Design}
We propose a structured construction of APSK constellations and prove that it achieves the constant gap to capacity regardless of $\mathrm{SNR}_{\text{c}}$. Later, we then build upon this foundation to propose a semi-analytical parametric APSK family that exhibits a tradeoff between asymptotically optimal communication and optimal sensing performance.
The following proposition specifies the explicit scaling laws for an asymptotically communication optimal APSK constellation, including the number of rings, ring radii, inter-ring spacing, and constellation points per ring.

\begin{proposition}
\label{pro1}
{\itshape
Let $\msf{X}$ be uniformly distributed over a normalized $2^m$-APSK constellation $\mathcal{X}$. The minimum Euclidean distance satisfies
$d_{\min}^2(\msf{X})
=
\Theta(2^{-m})$ if the following hold:

\begin{enumerate}
\item The number of rings satisfies
\begin{align}
K_m = \Theta(\sqrt{2^m}).
\label{Km_c}
\end{align}

\item Let $\widetilde{r}_k$ be the radius of the $k$-th ring before normalization which satisfy
\begin{subequations}\label{radii_c}
\begin{align}
&\widetilde{r}_k
= \Theta\!\left(\frac{k}{\sqrt{2^m}}\right),
\quad 1\le k \le K_m, \label{radii_c_a}\\
&\widetilde{r}_{k+1}-\widetilde{r}_k
= \Theta\!\left(\frac{1}{\sqrt{2^m}}\right),
\quad 1\le k \le K_m-1. \label{radii_c_b}
\end{align}
\end{subequations}

\item The number of constellation points per ring satisfies
\begin{align}
N_k=
\begin{cases}
    a_mk, & 1\le k \le K_m-1,\\
    2^m-\sum_{i=1}^{K_m-1}N_i, & k=K_m,
\end{cases}
\label{number_c}
\end{align}
where $a_m$ is set to satisfy $N_{K_m}=\Theta(K_m)>0.$

\item The phase offsets are aligned across rings, i.e.,
\begin{align}
    \phi_{k+1}-\phi_{k}=0, \qquad 1\le k \le K_m-1.
    \label{phase_c}
\end{align}
\end{enumerate}
}
\end{proposition}
\IEEEproof{
The proof is in Appendix C.\hfill$\blacksquare$
}

\begin{corollary}
    When $m=\Omega\left(\log_2(\mathrm{SNR}_{\mathrm{c}})\right)$, one can easily show that the APSK construction in Proposition~\ref{pro1} with $d_{\min}^2(\msf{X})=\Theta(2^{-m})$
achieves a constant gap to capacity using~\eqref{gap}.\hfill$\blacksquare$
\end{corollary}

\subsection{Design Criterion for Sensing}

To analyze the sensing performance, we study the average conditional CRB in
\eqref{eq:avg_CRB_exact}. 
Since energy fluctuations affect the average conditional CRB, the symbol-energy variance provides a tractable sensing-related metric.
The following lemma characterizes the dependence of the average conditional CRB on the symbol-energy variance.

\begin{lemma}
{\itshape
Considering L i.i.d. sensing symbols $\{x_\ell\}_{\ell=1}^{L}$, with $x_\ell$ being a realization of $\msf{X}$ for all $\ell$. Moreover, assume that $|x_{\ell}|^2\ge \delta>0$ for some positive constant $\delta$ and all $\ell$.
The average conditional CRB can be bounded as
\begin{equation}
    \overline{\mathrm{CRB}}_{h_{\text{s}}}
    \le
    \frac{\sigma_{\text{s}}^2}{P}
    \left[\frac{1}{L}
+
\frac{\mathrm{Var}(|\msf{X}|^2)}{L^2\delta}
    \right].
\label{eqcrb}
\end{equation}
}
\label{crb_var}
\end{lemma}
\IEEEproof{
The proof is in Appendix D.\hfill$\blacksquare$
}

\begin{remark}
{\itshape
Compared to CRB,
the symbol-energy variance provides deeper insight into sensing-oriented constellation design, as it implicitly captures the impact of the number of rings and the ring radii on the sensing performance. This metric, or the equivalent fourth-order statistic term, kurtosis, has been adopted as the sensing design metric in several works and shown to directly govern key sensing performance measures, such as ambiguity-function sidelobes~\cite{Du2024}, autocorrelation sidelobes~\cite{Fliu2025TIT}, and detection probability~\cite{Geiger2026}. Our design, which is also based on the symbol-energy variance, can potentially be applied to these sensing scenarios.
\hfill$\blacksquare$
}
\end{remark}

From~\eqref{eqcrb}, a smaller symbol-energy variance is beneficial for sensing. 
In particular, the best sensing performance is expected from constant-modulus signaling, as proved in~\cite[Eq. (66)]{Xiong2023}.
For the APSK family, this corresponds to the single-ring case, namely phase shift keying (PSK). 
Hence, PSK is sensing-optimal among the considered APSK constellations.

\subsection{Parametric APSK Signaling for the C\&S Tradeoff}

We now propose a semi-analytical parametric APSK family that achieves the tradeoff between the asymptotically communication-optimal point and sensing-optimal point. The APSK family is constructed as follows.
\begin{enumerate}
\item The number of rings satisfies
\begin{align}
K_m=\left\lfloor \sqrt{\frac{2^{m+1}}{\alpha_m}} \right\rfloor,
\label{Km}
\end{align}
where $\alpha_m \in \mathbb{N}$ is a design parameter.

\item Let $\widetilde{r}_k$ be the radius of the $k$-th ring before normalization, which satisfies
\begin{align}
\widetilde{r}_k=\frac{f(k)}{\sqrt{2^m}},
\qquad 1\le k \le K_m ,
\label{radii}
\end{align}
where $f(k)$ is a perturbation function that adjusts the ring radii, satisfying $f(k)=k+o(k)$, $f(k)>0$ for $1\le k \le K_m$, and
$\widetilde{r}_{k+1}>\widetilde{r}_k$ for $1\le k \le K_m-1$.

\item The phase offsets are aligned across rings, i.e.,
\begin{align}
\phi_k=0,
\qquad 1\le k \le K_m .
\label{phase}
\end{align}

\item The number of constellation points per ring satisfies
\begin{align}
N_k=
\begin{cases}
    \alpha_m k, & 1\le k \le K_m-1,\\
    2^m-\sum_{i=1}^{K_m-1}\alpha_m i, & k=K_m.
\end{cases}
\label{number}
\end{align}
\end{enumerate}

For this design, the minimum distance satisfies
\begin{equation}
    d_{\min}^2(\msf{X})=\Theta\!\left(\frac{1}{\alpha_m 2^m}\right).
    \label{dmindesign}
\end{equation}

When $\alpha_m$ is a constant, the family of APSK in~\eqref{Km}-\eqref{number}
reduces to a subset of the APSK construction in Proposition~\ref{pro1}. In contrast, when
$\alpha_m=2^m$, the APSK constellation degenerates into a single-ring PSK constellation,
which is sensing-optimal due to its constant-modulus property.

In the following remark, we discuss the parameters that dictate the C\&S tradeoff, considering a fixed constellation size.

\begin{remark}
    {\itshape
    According to the sensing-optimal design, one may think that an APSK constellation with fewer rings can have better sensing performance. However,
reducing the number of rings alone does not
guarantee a smaller symbol-energy variance. To make this explicit, consider two
normalized APSK constellations with $K_m<K_m'$ rings. The constellation with
$K_m$ rings has a smaller symbol-energy variance than that with $K_m'$ rings if
and only if $\sum_{k=1}^{K_m} N_k r_k^4
<
\sum_{k=1}^{K_m'} N_k' {r_k'}^4 .$
    Thus, when $K_m$ is reduced, $r_k$ should be designed such that the above fourth-moment condition is satisfied.
    In other words,
    the normalized ring radii must become more concentrated. This observation also clarifies the role of the perturbation design in~\eqref{radii}.
    
    On the other hand, increasing the number of rings up to the scaling in Proposition~\ref{pro1} improves the minimum distance as shown in~\eqref{dmindesign} and hence the mutual information by maintaining a constant gap to capacity. 
    Therefore, the number of rings and the energy allocation among rings jointly control the C\&S performance tradeoff, while an exact closed-form characterization remains challenging and is left for future work.
    }
    \hfill$\blacksquare$
\end{remark}

\section{Numerical Results}

In this section, we present simulation results to evaluate the parametric APSK family. We use the 
$\mathrm{Var}(|\msf{X}|^2)$-rate tradeoff as a proxy, since 
Lemma~\ref{crb_var} relates the average conditional CRB to the symbol-energy variance through an upper bound.

\begin{figure}[t]
    \centering
    \includegraphics[width=0.8\linewidth]{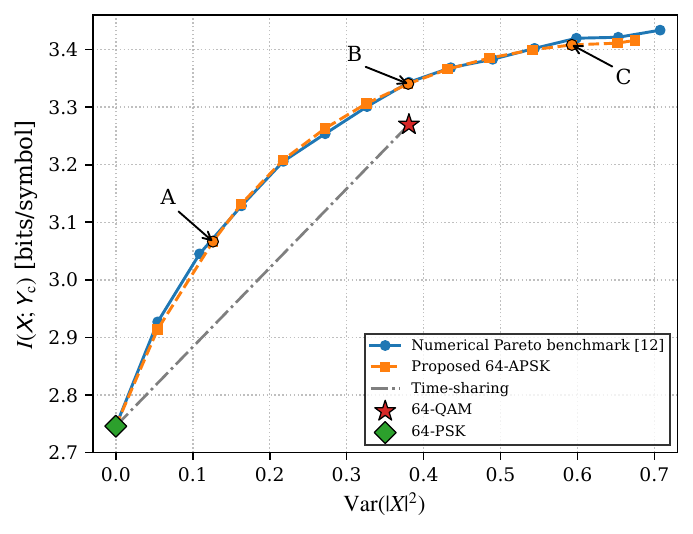}
    \caption{The achievable $\mathrm{Var}(|\msf{X}|^2)$-rate region at $\mathrm{SNR}_{\text{c}}=10~\mathrm{dB}$.}
    \label{10dB}
\end{figure}

\begin{figure}[t]
    \centering
    \includegraphics[width=0.8\linewidth]{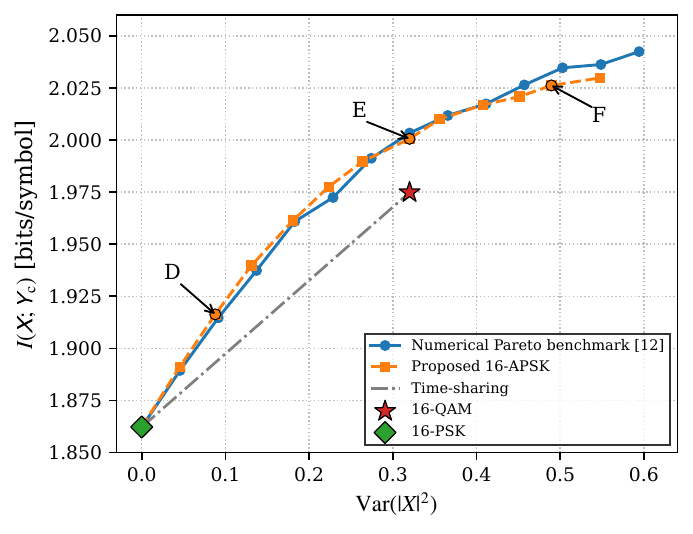}
    \caption{The achievable $\mathrm{Var}(|\msf{X}|^2)$-rate region at $\mathrm{SNR}_{\text{c}}=5~\mathrm{dB}$.}
    \label{5dB}
\end{figure}

We first consider $\mathrm{SNR}_{\text{c}}=10~\mathrm{dB}$ and
$m=6$. 
The number of rings is determined by~\eqref{Km}. 
For $\alpha_m\in[2,33]$, this yields $K_m\in[1,8]$. 
The perturbation function in~\eqref{radii} is chosen as $f(k)=k-c\sqrt{k}+b$, where $b$ and $c$ are the parameters
to maximize the mutual information for a given symbol-energy variance.
They are searched from 0 to 2 with a step size of 0.25.
The results for $\mathrm{SNR}_{\text{c}}=5~\mathrm{dB}$ are generated in a similar manner with $m=4$. 
The $\mathrm{Var}(|\msf{X}|^2)$-rate tradeoff achieved by the parametric APSK family, the Pareto boundary of $\mathrm{Var}(|\msf{X}|^2)$-rate region achieved by time-independent, circularly symmetric, and otherwise unconstrained continuous input distributions using MBA~\cite{Guo2026TWC}, and the time-sharing tradeoff region between PSK and quadrature amplitude modulation (QAM) are shown in Figs.~\ref{10dB} and~\ref{5dB}, respectively.
Table~1 lists the specific parameter settings corresponding to several selected APSK constellations in Figs.~\ref{10dB} and~\ref{5dB}.
In particular, points D, E, and F correspond to two-ring configurations, for which the tradeoff is primarily governed by radii.

\begin{table}[t]
    \centering
    \caption{Parameter settings corresponding to the labeled points.}
    \label{tab:labeled-points}
    \begin{tabular}{c c c c c}
        \hline
        Point & $\alpha_m$ & $b$ & $c$ & $K_m$\\
        \hline
        A & 16  & 0.50 & 0.00 & 2\\
        B & 5  & 0.50 & 0.75 & 5\\
        C & 5 & 1.25 & 2.00 & 5\\
        D & 4 & 1.00 & 0.75 & 2\\
        E & 5 & 0.50 & 1.25 & 2\\
        F & 8 & 0.25 & 0.75 & 2\\
        \hline
    \end{tabular}
\end{table}

Observe that the parametric APSK family achieves a tradeoff region very close to the Pareto boundary in~\cite{Guo2026TWC}. The minor apparent crossings of the numerically approximated boundary may be due to finite numerical resolution, convergence tolerance, and finite sampling of the tradeoff parameters in the MBA.
Our results also significantly outperform the time-sharing region between PSK and QAM. In addition, from Table~1, we see that the constellations closer to the sensing-optimal design tend to have fewer rings with more concentrated radii, whereas those closer to the communication-optimal design typically have more rings with more dispersed radii.

\section{Conclusion}
In this paper, we proposed a semi-analytical APSK design framework for ISAC.
We first proposed a semi-analytical asymptotically communication-optimal APSK construction whose key parameters are governed by explicit scaling laws and proved the constant-gap optimality. For sensing, we showed that the CRB 
can be upper-bounded in terms of the symbol-energy variance,
indicating that the optimal design is constant-modulus signaling.
To bridge the communication-optimal and sensing-optimal design, we proposed a semi-analytical parametric APSK family where the C\&S tradeoff can be controlled by both the number of rings and the energy allocation among rings. Simulation results show that the C\&S tradeoff achieved by the APSK family lies very close to the Pareto boundary obtained by the MBA with continuous inputs~\cite{Guo2026TWC}, and is superior to that of the PSK-QAM time-sharing baseline.

\section*{Appendix A}
\begin{lemma}[Lemma~5 of~\cite{Qiu2021TIT}]
{\itshape
Let $\msf{X}$ be a discrete random variable uniformly distributed over a two-dimensional constellation $\Lambda$ with minimum distance $d_{\min}(\Lambda) > 0$. Let $\msf{Z} \sim \mathcal{CN}(0,1)$ and independent of $X$. Then
\begin{equation}
I(\msf{X}; \msf{X}+\msf{Z}) \ge H(\msf{X}) - \log_2 \left( 2\pi e \left( \frac{4}{\pi d_{\min}^2(\Lambda)} + \frac{1}{4} \right) \right).
\end{equation}
}
\IEEEproof{The proof is in Appendix~B of~\cite{Qiu2021TIT}.}
\hfill$\blacksquare$
\label{mutualboundlemma}
\end{lemma}

\section*{Appendix B\\Proof of Lemma~\ref{dminenergy}}
The constellation is normalized such that 
$\frac{1}{2^m}\sum_{x\in\mathcal{X}} |x|^2 = 1.$
Hence, no more than $2^{m-1}$ points can satisfy $|x| > \sqrt{2}$; otherwise,
\begin{align}
    \sum_{x\in\mathcal{X}} |x|^2 > 2^{m-1}\cdot (\sqrt{2})^2 = 2^m,
\end{align}
which contradicts the unit average energy constraint.

Therefore, at least $2^{m-1}$ points lie in the closed disk
\begin{align}
    \mathcal{D}(0,\sqrt{2}) = \{ z \in \mathbb{C} : |z| \le \sqrt{2} \}.
\end{align}
Denote these points by $y_1,\dots,y_{2^{m-1}}$.

Now, construct disks centered at the points $y_i$ with radius $d_{\min}(\msf{X})/2$, i.e., $\mathcal{D}\left(y_i,\frac{d_{\min}(\msf{X})}{2}\right).$
Since the minimum distance between any two constellation points is at least $d_{\min}(\msf{X})$, these disks are pairwise disjoint. Moreover, all these disks are contained in the larger disk $\mathcal{D}\left(0,\sqrt{2}+\frac{d_{\min}(\msf{X})}{2}\right).$

By comparing the total area, we obtain
\begin{align}
2^{m-1}\cdot \pi \left(\frac{d_{\min}(\msf{X})}{2}\right)^2
\le
\pi \left(\sqrt{2}+\frac{d_{\min}(\msf{X})}{2}\right)^2.
\end{align}
This implies that $d_{\min}^2(\msf{X}) \le \frac{96+64\sqrt{2}}{2^m}$ for $m\ge2$, which completes the proof.

\section*{Appendix C\\Proof of Proposition~\ref{pro1}}

From~\eqref{radii_c_a} and~\eqref{number_c}, we have
$N_k\widetilde{r}_k^2
=\Theta(k)\Theta(k^2/2^m)=\Theta(k^3/2^m)$.
Substituting this into the pre-normalization average symbol energy gives
\begin{align}
E_0
=
\frac{1}{2^m}\sum_{k=1}^{K_m}N_k\widetilde{r}_k^2
=
\Theta\!\left(\frac{1}{2^m}\sum_{k=1}^{K_m}\frac{k^3}{2^m}\right)
=
\Theta\!\left(\frac{K_m^4}{2^{2m}}\right)
=
\Theta(1),
\end{align}
where the last equality follows from~\eqref{Km_c}.Hence, normalization only introduces a constant scaling factor, and~\eqref{radii_c} gives
\begin{subequations}\label{appendixc}
\begin{align}
    &r_k
    = \Theta\!\left(\frac{k}{\sqrt{2^m}}\right), \label{28}\\
    &r_{k+1}-r_k
    = \Theta\!\left(\frac{1}{\sqrt{2^m}}\right). \label{29}
\end{align}
\end{subequations}

For the intra-ring distance, by substituting~\eqref{28} and \eqref{number_c} into~\eqref{dminintra}, we obtain
\begin{align}
d_k^{\mathrm{intra}}
=
2\Theta\!\left(\frac{k}{\sqrt{2^m}}\right)
\Theta\!\left(\frac{1}{k}\right)
=
\Theta(\sqrt{2^{-m}}).
\end{align}
Together with the inter-ring distance in~\eqref{29} and phase-alignment
condition in~\eqref{phase_c}, this yields 
\begin{align}
   d_{k,k+1}^{\mathrm{inter}}
=
\Theta(\sqrt{2^{-m}}). 
\end{align}
Therefore, $d_{\min}(\msf{X})=\Theta(\sqrt{2^{-m}})$, and hence
\begin{align}
d_{\min}^2(\msf{X})
=
\Theta(2^{-m}).
\end{align}
This completes the proof.

\section*{Appendix D\\Proof of Lemma~\ref{crb_var}}
Since $|x_{\ell}|^2\ge \delta>0$ for all $\ell$,
$\sum_{\ell=1}^{L}|x_{\ell}|^2\ge L\delta$. Using the identity
\begin{align}
\frac{1}{\sum_{\ell=1}^{L}|x_\ell|^2}
=
\frac{1}{L}
-
\frac{\sum_{\ell=1}^{L}|x_\ell|^2-L}{L^2}
+
\frac{
\left(\sum_{\ell=1}^{L}|x_\ell|^2-L\right)^2
}{
L^2\sum_{\ell=1}^{L}|x_\ell|^2
},
\end{align}
we have
\begin{align}
&\mathbb E\!\left[
\frac{1}{\sum_{\ell=1}^{L}|x_\ell|^2}
\right]
=
\frac{1}{L}
+
\mathbb E\!\left[
\frac{
\left(\sum_{\ell=1}^{L}|x_\ell|^2-L\right)^2
}{
L^2\sum_{\ell=1}^{L}|x_\ell|^2
}
\right]  \notag\\
&\le
\frac{1}{L}
+
\frac{
\mathrm{Var}\!\left(\sum_{\ell=1}^{L}|x_\ell|^2\right)
}{
L^3\delta
} 
=
\frac{1}{L}
+
\frac{\mathrm{Var}(|\msf{X}|^2)}{L^2\delta}.
\end{align}

\newpage
\bibliographystyle{IEEEtran}
\bibliography{ref}

\end{document}